\documentstyle[aps]{revtex}

\begin{document}
\title{Accessible information and quantum operations }
\author{Qing-yu Cai}
\address{Laboratory of Magentic Resonance and Atom and Molecular Physics, Wuhan\\
Institute of Mathematics, The Chinese Academy of Science, Wuhan,\\
430071, People's Republic of China}
\maketitle

\begin{abstract}
The accessible information decreases under quantum operations. We analyzed
the connection between quantum operations and accessible information. We
show that a general quantum process cannot be operated accurately.
Futhermore, an unknown state of a closed quantum system can not be operated
arbitrarily by a unitary quantum operation.

\begin{itemize}
\begin{description}
\item  PACS number: 03.65.-w, 03.65.Bz, 03.67.-a,
\end{description}
\end{itemize}
\end{abstract}

It is impossible to perfectly clone an unknown quantum state using unitary
evolution [1]. That non-orthogonal states can't be distinguished is well
known [2]. In [3], Mor and Terno showed that an arbitrary state cannot be
disentangled, a universal disentangling machine cannot exist. It is well
known that the laws of quantum mechanics allow us to do many things superior
comparing with classical mechanics. At the same time, quantum physics
establishes a set of negative rules stating thing that cannot be done.

In this paper, we shall prove that a general quantum process can not be
operated accurately unless which can be performed by a unitary operation.
The key of our proof is the theorem which was conjectured by Gordon in 1964
[4] and proved by Holevo in 1973 [5]. Suppose X is a quantum system which is
prepared in the mixed state $\rho _{x}$ with a with a probability $p_{x}$: 
\begin{equation}
\rho =\sum_{x}p_{x}\rho _{x}.
\end{equation}
A quantum measurement on the state has been given to identify X on the
measurement result Y. A good measure of how much information has been gained
about X from the measurement is the mutual information H(X : Y) between X
and the measurement outcome Y. The mutual information H(X : Y) of X and Y
measures how much information X and Y have in common. Holevo's theorem
states that 
\begin{equation}
H(X:Y)\leq S(\rho )-\sum_{x}p_{x}S(\rho _{x}),
\end{equation}
where S($\rho $)=-Tr($\rho \log \rho $), the von Neumann entropy of the
density operator $\rho $. The quantity appearing on the right hand side of
the inequality 
\begin{equation}
\chi =S(\rho )-\sum_{x}p_{x}S(\rho _{x})
\end{equation}
is an upper bound on the accessible information.

First, in this paper, we aim to show that the Holevo chi quantity decreases
under quantum operations: 
\begin{equation}
\chi ^{\prime }=S(\varepsilon (\rho ))-\sum_{x}p_{x}S(\varepsilon (\rho
_{x}))\leq \chi =S(\rho )-\sum_{x}p_{x}S(\rho _{x}),
\end{equation}
where $\varepsilon $ is a quantum operation. Suppose we want to perform a
quantum process, which can be realized by a quantum operation, $\varepsilon $%
. After a quantum process, we find the mutual information decreased, in
another world, we find that the information we can gain from the quantum
system is less than previous. We know the uncertainty of this quantum states
increases because of the decreasing of the accessible information. That is,
the result after the quantum operation is unreliable. Also, it tell us this
quantum process can not be operated accurately (If this quantum can be
operated accurately, the accessible information would not decrease.).

It is well known that the dynamics of a closed quantum system are described
by a unitary transform. A natural way to describe the dynamics of an open
quantum system is to regard it as arising from an interaction between the
system of interest, which we shall call $principal$ $system$, and an $%
environment$, which together form a closed quantum, Fig.1. [6]. Suppose we
have a system in state $\rho $. In general, the final state of the system, $%
\varepsilon (\rho )$, may not be a unitary transformation to the initial
state $\rho $. After the transformation $U$ the system no longer interacts
with the environment, and thus we perform a partial trace over the
environment to obtain the reduced state of the system alone: 
\begin{equation}
\varepsilon (\rho )=tr_{E}[U(\rho \otimes \rho ^{E})U^{\dagger }]
\end{equation}
Using the Stinespring dilation theorem [7], we know any possible
transformation $\rho \rightarrow \varepsilon (\rho )$, can be described, if
the $principical$ $system$ has a Hilbert space of $d$ dimensions, then it
suffices to model the environment as being in a Hilbert space of no more
than $d^{2}$ dimensions. We suppose that the quantum system X is prepared in
the state $\rho _{x}^{X}$ with a probability $p_{x}$ initially. The state of
environment E is $\rho _{0}^{E}$, independent of the quantum system X. The
initial state that includes the quantum system and environment we obtain is 
\begin{equation}
\rho ^{EX}=\sum_{x}p_{x}\rho _{x}^{EX},
\end{equation}
where 
\begin{equation}
\rho _{x}^{EX}=\rho _{0}^{E}\otimes \rho _{x}^{X}
\end{equation}
Because $\rho _{0}^{E}$ is independent to every $x$, so we obtain 
\begin{equation}
\chi ^{EX}=\chi ^{X}.
\end{equation}

A quantum process of system X now take place. It can be realized by a
unitary operation on the states $\rho ^{EX}$: 
\begin{equation}
\stackrel{\sim }{\rho }^{EX}=U(\rho ^{EX})U^{\dagger }
\end{equation}
After the unitary operation $U$ the system no longer interacts with the
environment, and thus we perform a partial trace to over the environment to
obtain the reduce state of the system alone: 
\begin{equation}
\stackrel{\sim }{\rho }^{X}=tr_{E}[U(\rho \otimes \rho ^{E})U^{\dagger }].
\end{equation}
In [8], it has been shown that $\chi $ is nonincreasing under the partial
trace operation. Suppose $|x_{i}^{X}><x_{i}^{X}|$ are the states of a system 
$X$, where states $|x_{i}^{X}>$ are orthogonal. Introduce an auxiliary
system $Y$. Define a joint state of $XY$ by: 
\begin{equation}
\rho ^{XY}=\sum_{i}p_{i}|x_{i}^{X}><x_{i}^{X}|\otimes \rho _{i}^{Y}.
\end{equation}
The the entropy of the joint state is 
\begin{equation}
S(\rho ^{XY})=H(p_{i})+\sum_{i}p_{i}S(\rho _{i}^{Y}).
\end{equation}
$H(p_{i})$ is Shannon entropy as a function of a probability distribution.

Consider the state 
\begin{equation}
\rho ^{XYZ}=\sum_{i}p_{i}\rho ^{XY}\otimes |i^{Z}><i^{Z}|,
\end{equation}
Where $|i^{Z}>$ is an orthogonal set of states: 
\begin{equation}
S(\rho ^{XYZ})=H(p_{i})+\sum_{i}p_{i}S(\rho _{i}^{XY})
\end{equation}
\begin{equation}
S(\rho ^{XZ})=H(p_{i})+\sum_{i}p_{i}S(\rho _{i}^{X}).
\end{equation}
Using strong subadditivity [9] which is a property of the entropy functional
for a trio of quantum systems, $X$, $Y$, $Z$, 
\begin{equation}
S(\rho ^{XYZ})+S(\rho ^{Y})\leq S(\rho ^{XY})+S(\rho ^{YZ}),
\end{equation}
we can obtain 
\begin{equation}
S(\rho ^{X})-\sum_{i}p_{i}S(\rho _{i}^{X})\leq S(\rho
^{XY})-\sum_{i}p_{i}S(\rho _{i}^{XY}).
\end{equation}
That is 
\begin{equation}
\chi ^{X}\leq \chi ^{XY}.
\end{equation}
$\chi $ is nonincreasing under the partial trace operation.

Suppose now we have a quantum system X. $\varepsilon $ is a general quantum
operation. Let perform the quantum operation act on the system 
\begin{equation}
\stackrel{\sim }{\rho }^{X}=\varepsilon (\rho ^{X})
\end{equation}
From Eq.(5), we know this process can be realized be a unitary operation $U$
on the system and environment $\rho ^{XE}$: 
\begin{equation}
\stackrel{\sim }{\rho }^{XE}=U(\rho \otimes \rho ^{E})U^{\dagger }.
\end{equation}
\begin{equation}
\stackrel{\sim }{\rho }^{X}=tr_{E}(\stackrel{\sim }{\rho }^{XE})
\end{equation}
Since unitary operation does not change entropy of quantum, that is $%
\stackrel{\sim }{\chi }^{XE}=\chi ^{XE}$. Considering Eq.(8) and Eq.(18), we
can draw a conclusion: 
\begin{equation}
\stackrel{\sim }{\chi }^{X}\leq \chi ^{X}.
\end{equation}
That is, the Holevo chi quantity decreases under quantum operations. From
Eq.(2) we know, the Holevo chi quantity is an upper bound on the accessible
information. After a quantum operation, the Holevo chi quantity decreases,
which means some of the information we can get (theoretic) before this
quantum operation has lost. Since the information we can get of the system
decreases, the uncertainty of final state increases. For an unknown state,
the Holveo chi quantity decrease under quantum operations tell us we cannot
operate it accurately because of the information we can obtain about the
state decreasing.

Consider the likelihood: 
\begin{equation}
\stackrel{\sim }{\chi }^{X}=\chi ^{X}.
\end{equation}
In Eq.(5), of course, if $U$ does not involve any interaction with the
environment, then $\varepsilon (\rho )=\stackrel{\sim }{U}\rho \stackrel{%
\sim }{U}^{\dagger }$, where $\stackrel{\sim }{U}$ is the part of $U$ which
acts on the system alone. It is clear that $\stackrel{\sim }{\chi }^{X}=\chi
^{X}$. That is, when we want to operate an unknown quantum state, if this
process can be performed by a unitary operation, we can say that this
process can be operated accurately because the upper bound on the accessible
information does not decrease. When $U$ involve interaction with the
environment, it would result in that information exchanges between the
quantum system and environment. The accessible information of the quantum
system must decrease. In a general way, a quantum process can not be
operated accurately. We will show this by a common quantum process below.

Consider a simple but common quantum process. Let suppose two subsystems $A$
and $B$. $A$ starts out in an unknown but pure quantum state, $|\varphi >$. $%
B$ starts in some standard pure state, $|s>$. The quantum process can be
described like this: 
\begin{equation}
U|\varphi >|s>=|\varphi >|f(\varphi ,s)>,
\end{equation}
where $U$ is a unitary evolution, Fig.2. Suppose this procedure works for
two particular states, $|\varphi >$ and $|\phi >$. Then we have 
\begin{eqnarray}
U|\varphi &>&|s>=|\varphi >|f(\varphi ,s)>, \\
U|\phi &>&|s>=|\phi >|f(\phi ,s)>.
\end{eqnarray}
Taking the inner product of this two equations gives 
\begin{eqnarray}
&<&\varphi |\phi >=<\varphi |\phi ><f(\varphi ,s)|f(\phi ,s)> \\
&<&\varphi |\phi >(1-<f(\varphi ,s)|f(\phi ,s)>)=0.
\end{eqnarray}
It has only two solutions, either $|\varphi >$and $|\phi >$ are orthogonal
or $|f(\varphi ,s)>=|f(\phi ,s)>$. Only those states which are orthogonal to
one another can be evolved, or else, the standard state of subsystem $B$
would be invariable. Therefore, a general quantum device described by
Eq.(24) is not possible (Reasonably, we can consider the two subsystems as
one joint closed system. Therefore, we know the states of a closed quantum
system can not be operated accurately in general.). It can be explained like
this: If this process can be realized. That is to say, the state of system $%
A $ does not change. The information of system $A$ does not decrease. We can
repeat this quantum process to obtain more and more states $|f(\varphi ,s)>$%
. Then we can draw some information about system $A$ from system $B$ which
now is in the state $|f(\varphi ,s)>$. The information about system $A$
would increase. It is impossible! When $|f(\varphi ,s)>=|f(\phi ,s)>$, the
states of subsystem $B$ is constant. No information exchanges between the
two subsystems $A$ and $B$. The quantum process described by Eq.(24) does
not increase the accessible information. It seems like a $controlled-U$
process: the subsystem $B$ as the control qubit, and the subsystem $A$ as
the target qubit.

Consider the unitary operation $U$ is $controlled$ $operation$, with the
subsystem $A$ the control qubit, and the subsystem $B$ initially as the
target qubit, Fig.3. '$If$ $A$ $is$ $ture,$ $then$ $do$ $B$': 
\begin{equation}
U|\gamma >|\psi >=|\gamma >|f(\gamma ,\psi )>.
\end{equation}
This type of $controlled$ $operation$ is one of the most useful in quantum
computing. The proof of universality for $controlled-NOT$ gate has been
given in [10]. The reason why the $controlled$ $operation$ is universal is
that we have full information about the control qubit. The target qubit is
in an arbitrary pure state. Clearly, in this process, the accessible
information of control and target qubit does not increase.

Let $|\varphi >=|f(\varphi ,s)>$, Eq.(24) will becomes a copying procedure: 
\begin{equation}
U|\varphi >|s>=|\varphi >|\varphi >.
\end{equation}
Clearly, a universal cloning device is impossible. This is the very famous
no-cloning theorem [1]. The connection between the accessible information
and a copying procedure is more obvious. Suppose Alice prepares one of two
non-orthogonal quantum states $|\varphi >$ and $|\phi >$ with respective
probabilities $p$ and $1-p$. If the law of quantum mechanics allowed Bob to
get full information by measurement to identify which of the two states $%
|\varphi >$ and $|\phi >$ Alice had prepared, then he could clone the states
simply. On the other hand, if Bob can clone the two states, then he could
repeatedly apply the clone device to get a large number of copies. Then he
could get full information by measurement. But, the accessible information
decreases under quantum operations, which forbid a universal cloning.

Consider a disentanglement process: 
\begin{equation}
\varepsilon (\rho _{12})\rightarrow tr_{2}(\rho _{12})\otimes tr_{1}(\rho
_{12}),
\end{equation}
where $\rho _{12}$ is a pure state of two subsystem.. It has been showed
that an arbitrary state cannot be disentangled, by a physical allowable
process, into a tensor product of its reduced density matrices [3]. Consider 
$\rho _{12}$ is a pure state, then the Holevo chi quantity $\chi (\rho
_{12})=0$. After the disentangling process, it has that the Holevo chi
quantity $\chi (tr_{2}(\rho _{12})\otimes tr_{1}(\rho _{12}))\geq 0$. We can
prove this by using concavity of the entropy: 
\begin{equation}
S(\sum_{i}p_{i}\rho _{i})\geq \sum_{i}p_{i}S(\rho _{i}).
\end{equation}
After the disentangling quantum states process, that is 
\begin{eqnarray}
\chi (tr_{2}(\rho _{12})\otimes tr_{1}(\rho _{12})) &=&S(\sum_{i}p_{i}\rho
_{i})-\sum_{i}p_{i}S(\rho _{i})\geq 0 \\
\chi (tr_{2}(\rho _{12})\otimes tr_{1}(\rho _{12})) &\geq &\chi (\rho _{12})
\end{eqnarray}
Since the equality holds if and only if all the states $\rho _{i}$ for which 
$p_{i}>0$ are identical, we know that a general disentangling quantum states
would necessarily increase the Holevo chi quantity. Thus, it tell us a
universal disentangling machine cannot exist.

In summary, we analyzed the connection between quantum operations and
accessible information. For an unknown state, generally, it cannot be
operated accurately because the Holevo chi quantity decrease under quantum
operations. At the same time, quantum mechanics law shows that
non-orthogonal states cannot be reliably distinguished. Generally, for an
unknown state, we neither can get full information about it, nor can we
operated it accurately. This, also, is the restriction that the law of
quantum mechanics gives.

\subsection{References:}

[1]. D. Dieks. Phys. Lett. A, 92, 271-272, (1982); W. K. Wootters and W. H.
Zurek. Nature, 299, 802-803, (1982); H. Barnum, C. M. Caves, C. A. Fuchs, R.
Jozsa, and B. Schumacher. Phys. Rev. Lett., 76, 2818-2821, (1996); T. Mor.
Phys. Rev. Lett., 80, 3137-3140, (1998).

[2]. A. Peres. Phys.Lett. A, 128, 19, (1988); L.-M. Duan and G.-C. Guo,
Phys. rev. Lett., 80, 4999-5002, (1998)

[3]. T. Mor. Phys. Rev. Lett., 83, 1451-1454, (1999); D. R. Terno. Phys.
Rev. A, 59, 3320-3324, (1999).

[4]. J. P. Gordon. Noise at optical frequencies; information theory. In P.
A. Miles, editor, $quantum$ $Electronics$ $and$ $Coherent$ $Light$,
Proceeding of International School of Physics 'Enrico Fermi' XXXI, Academic
Press, New York, 1964.

[5]. A. S. Holevo. Probl. Peredachi Inf. 9, 3 (1973) [Prob. Inf. Transm.
(USSR) 9, 177 (1973)].

[6]. M. A. Nielsen and I. L. Chuang. quantum Computation and Quantum
Information (Cambridge University Press, Cambridge, UK, 2000)

[7]. W. F. Stinespring, Proc. Am. Math. Soc. 6, 211 (1955).

[8]. B. Schumacher, M. Westmoreland, and W. K. Wootters. Phys. Rev. Lett.,
76, 3452-3455, (1996)

[9]. E. H. Lieb and M. B. Ruskai. Phys. Rev. Lett., 30, 434-436, (1973); E.
H. Lieb and M. B. Ruskai. J. Math. Phys., 14, 1938-1941, (1973).

[10]. A. Barenco, C. H. Bennett, R. Cleve, D. P. DiVincenzo, N. Margolus, P.
shor, T. Sleator, John A. Smolin, and H. Weinfurter. Phys. Rev. A, 52,
3457-3467, (1995).

\end{document}